\documentclass[12pt,preprint]{aastex}

\begin{document}

\title{{\it Chandra} X-ray Detection of the Enigmatic
  Field Star BP Psc}

\author{Joel H. Kastner\altaffilmark{1}, Rodolfo Montez
  Jr.\altaffilmark{1}, David Rodriguez\altaffilmark{2}, 
  Nicolas Grosso\altaffilmark{3}, 
  B. Zuckerman\altaffilmark{2}, Marshall D. Perrin\altaffilmark{2},
  Thierry Forveille\altaffilmark{4},
  James R. Graham\altaffilmark{5}}

\altaffiltext{1}{Center for Imaging Science, Rochester Institute of
  Technology, 54 Lomb Memorial Drive, Rochester NY 14623
  (jhk@cis.rit.edu)} 
\altaffiltext{2}{Dept.\ of Physics \& Astronomy,
  University of California, Los Angeles 90095, USA}
\altaffiltext{3}{Observatoire astronomique de Strasbourg, Universit\'e
  de Strasbourg, CNRS, UMR 7550, 11 rue de l’Université, F-67000
  Strasbourg, France} 
\altaffiltext{4}{Laboratoire d'Astrophysique de
  Grenoble, Universit\'e Joseph Fourier --- CNRS, BP 53, 38041
  Grenoble Cedex, France} 
\altaffiltext{5}{Astronomy Dept., 601
  Campbell Hall, University of California, Berkeley, CA 94720}

\begin{abstract}
  BP Psc is a remarkable emission-line field star that is orbited by a
  dusty disk and drives a parsec-scale system of jets. We report the
  detection by the {\it Chandra} X-ray Observatory of a weak X-ray
  point source coincident with the centroids of optical/IR and
  submillimeter continuum emission at BP Psc. As the star's
  photosphere is obscured throughout the visible and near-infrared,
  the {\it Chandra} X-ray source likely represents the first detection
  of BP Psc itself. The X-rays most likely originate with magnetic
  activity at BP Psc and hence can be attributed either to a stellar
  corona or to star-disk interactions. The log of the ratio of X-ray to
  bolometric luminosity, $\log{(L_X/L_{bol})}$, lies in the range $-5.8$
  to $-4.2$. This is smaller than $\log{(L_X/L_{bol})}$ ratios typical of
  low-mass, pre-main sequence stars, but is well within the
  $\log{(L_X/L_{bol})}$ range observed for rapidly-rotating (FK Com-type) G
  giant stars. Hence, the {\it Chandra} results favor an exotic model
  wherein the disk/jet system of BP Psc is the result of its very
  recently engulfing a companion star or giant planet, as the primary
  star ascended the giant branch.
\end{abstract}

\section{Introduction}

Since its discovery by Stephenson (1986), the remarkable field star BP
Psc (= StH$\alpha$ 202) has been classified as a classical T Tauri
star (cTTS) on the basis of its strong H$\alpha$ and forbidden line
emission. The discovery by Zuckerman et al.\ (2008; hereafter ZMS08) of
an orbiting, dusty circumstellar disk and an enormous (pc-scale)
system of highly collimated outflows (jets) would appear, at first
glance, to provide strong support for such a cTTS
classification. However, as noted by ZMS08,
it is not at all clear that BP Psc is a %disk-enshrouded, jet-driving,
pre-main sequence (pre-MS) star.

First, although its jet system resembles those of cloud-embedded young
stellar objects, BP Psc is located at high galactic latitude, far from
any known star forming cloud. Furthermore, unlike the rare (though
increasingly scrutinized) ``isolated,'' cTTS systems within $\sim100$
pc (Torres et al.\ 2008) --- the best-known examples being TW Hya
(Kastner et al.\ 1997; Zuckerman \& Song 2004; and references therein)
and V4046 Sgr (Kastner et al.\ 2008b; Rodriguez et al.\ 2010; and
references therein) --- BP Psc is not associated with any known sparse
group of young stars (ZMS08). Second, its photospheric Li abundance
appears to be anomalously weak for a K-type pre-MS star of age $<100$
Myr, and its gravity-sensitive photospheric absorption lines also call
into question such a classification (ZMS08).  Third, the mm-wave  molecular
spectrum of its circumstellar disk --- specifically, its weak CN and
HCO$^+$ emission --- is atypical of the disks of low-mass, pre-MS
stars (Kastner et al.\ 2008a).

In light of the first two considerations, ZMS08
proposed that BP Psc is most likely a first-ascent giant at a
distance $\sim300$ pc. They speculated that the BP Psc disk/jet system
might therefore be the result of a recent, catastrophic interaction in
which a low-mass companion star (or perhaps giant planet) had been
consumed by BP Psc. Though somewhat exotic, a similar (companion
engulfment) scenario may also apply to the first-ascent giant star TYC
4144 329 2 (Melis et al.\ 2009): the post-MS nature of TYC 4144 329 2
is well-established, yet (like BP Psc) this star is evidently actively
accreting gas from a dusty circumstellar disk.

Low-mass, pre-MS stars are highly luminous X-ray sources, with
typical (0.3--8.0 keV) X-ray luminosities 
$\log{L_X} \, {\rm [erg \; s}^{-1}] \stackrel{>}{\sim} 29.5$ and X-ray to
bolometric luminosity ratios in the range $\log{(L_X/L_{bol})} \sim -4$
to $-3$ (e.g., Preibisch \& Feigelson 2005; G\"{u}del et al.\ 2007a).
Hence, X-ray observations are diagnostic of stellar
youth. Furthermore, in rare cases, TTS jet systems may yield
shocks energetic enough to produce X-rays, yielding insight into
physical conditions within the jets (e.g., Kastner et al.\ 2005;
Guedel et al.\ 2005, 2007b).  Here, we report the results of {\it
  Chandra} X-ray Observatory observations of BP Psc intended to
determine whether BP Psc and/or its jet system emit X-rays and,
thereby, to better understand the nature of this unusual system.

\section{Observations}

{\it Chandra} observed BP Psc in 2009 January, with the Advanced CCD
Imaging Spectrometer (ACIS) as the detector. Due to spacecraft thermal
constraints, the observation was obtained in two pieces of duration
10.15 and 65.4 ks (OBSIDs 8900 and 10856, respectively, on 2009 Jan.\
12 and 13), for a total exposure time of 75.5 ks. Data were obtained
with ACIS CCDs S1, S2, S3, S4, I2, and I3 active. Each CCD has a
$\sim8'\times8'$ field of view with 0.49$''$ pixels, and is sensitive
to photons in the energy range 0.5--8.0 keV (with some additional
sensitivity down to 0.3 keV, for back-illuminated CCD S3). The
telescope was pointed such that BP Psc was positioned at the nominal
aimpoint of CCD S3, $\sim2'$ from the S2-S3 chip boundary. The merged
event data from the two exposures were subject to standard processing,
filtering, and calibration (via {\it Chandra} X-ray Center (CXC) ACIS
pipeline v7.6.1 and CALDB v3.5.1).

%\subsection{Source detection and astrometry}

We ran the CXC (CIAO\footnote{http://cxc.harvard.edu/ciao/}) source
detection tool \verb+celldetect+ on the filtered event list,
restricted to CCDs S2 and S3 (the other active CCDs lie far off the
best focal surface). We find 26 sources, including BP Psc itself (\S
3.1), were detected at a significance $\ge 3\sigma$ (using Poisson
statistics) on these two CCDs (Fig.~\ref{fig:BPPscImage}, right).
Four of these S2 and S3 sources lie within $\sim1''$ of DENIS near-IR
sources; two of the four (BP Psc itself and an X-ray source detected on
CCD S2; \S 3.3) are also 2MASS sources. Based on the $\sim10$ X-ray
sources that lie within $\sim1''$ of Guide Star Catalog stars (not
including the BP Psc source), we estimate the absolute astrometric
accuracy of the two-part {\it Chandra} pointing to be $\sim0.2''$.

\section{Results}

\subsection{The X-ray source at BP Psc}

In Fig.~\ref{fig:BPPscImage} we display the wide-field H$\alpha$ image
originally published in ZMS08 alongside the broad-band (0.5--8.0 keV)
{\it Chandra}/ACIS image of the same field surrounding BP Psc. The
comparison reveals that {\it Chandra} has detected a faint, point-like
X-ray source at the position of BP Psc. Source detection indicates a
background-subtracted count rate of 0.24$\pm$0.07 ks$^{-1}$ for this
source, or $\sim18$ source photons detected during the $\sim75$ ks
exposure. The BP Psc X-ray source lies at J2000 coordinates
$\alpha=$23:22:24.72, $\delta=-$02:13:41.66, with a positional
uncertainty that is dominated by the estimated 
astrometric accuracy ($\pm$0.2$''$). Given the uncertainties, this
X-ray position is precisely coincident with that of the 2MASS position
of BP Psc ($\alpha=$23:22:24.70, $\delta=-$02:13:41.40) as well as
with the centroid of submillimeter continuum emission from BP Psc
($\alpha=$23:22:24.72$\pm$0.02, $\delta=-$02:13:41.55$\pm$0.2;
ZMS08). Analysis of photon arrival times (via a Kuiper test against a
Poisson process; Press et al.\ 1992) yields no evidence that the BP
Psc source is variable, although we are unable to rule out source
variability.

\begin{figure}[htb]
  \centering
\includegraphics[width=3in,angle=-90]{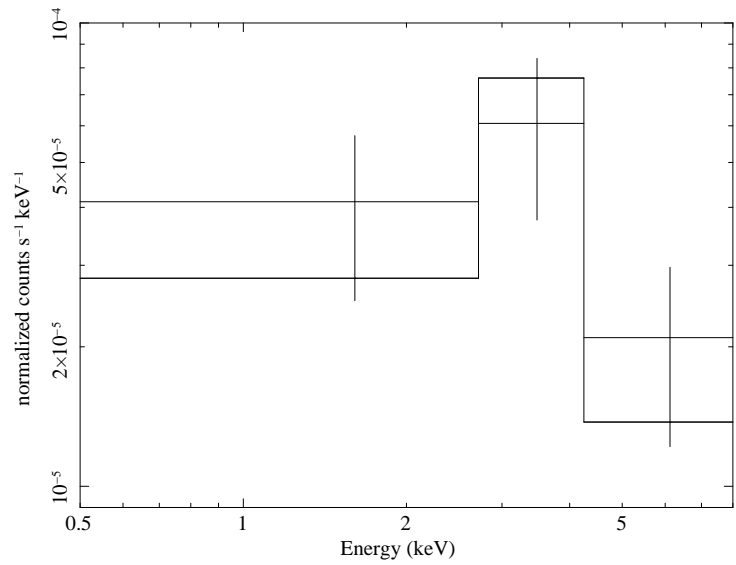}
\caption{Spectral energy distribution of the X-ray source at BP Psc
  (crosses) overlaid with the ``best model family'' spectral model
  (histogram; see Getman et al.\ 2010). The parameters describing this
  (two-component thermal plasma) model are $kT_1 = 0.8$ keV and
  $kT_2=1.8$ keV, with emission measure ratio EM$_1$/EM$_2$$=1.0$,
  metal abundance $Z=0.3$ (relative to solar), and intervening
  absorbing column $\log{N_H} \, [{\rm cm}^{-2}] = 22.9$.}
%> I have used AE to group the spectrum in 3 spectral bins (which
%> corresponds here to SNR=1.6; yes this source is really faint...),
%> and I have overlaid the best model family (logLhc=28.5, kT1=0.8 keV,
%> kT2=1.8, EM2/EM1=1.0, Z=0.3).
\label{fig:BPPsc_spec} 
\end{figure}

Given the small number of photons detected, it is not reasonable to
attempt to deduce the characteristic source temperature $T_X$ and
line-of-sight absorbing column $N_H$ and, hence, intrinsic source flux
$F_X$ via standard X-ray spectral analysis. 
%This then severely limits
%our ability to estimate the intrinsic source X-ray flux $F_X$. 
However, useful constraints on $N_H$ and $F_X$ can be obtained via the
median energy of the detected source photons. Specifically, Feigelson
et al.\ (2005) demonstrated that Orion Nebula Cluster (ONC) pre-MS
star X-ray sources detected by {\it Chandra}/ACIS-I display a
relatively tight correlation between median X-ray energy and
$N_H$. Getman et al.\ (2010) refined and further generalized the
Feigelson et al.\ (2005) results, demonstrating (via spectral
simulations) that the relationship between median X-ray energy and
$N_H$ should in principle apply to any source that can be modeled in
terms of two-temperature optically thin thermal plasma emission
suffering intervening absorption, and is applicable over a wide range
of intrinsic source luminosity (i.e., for ${L_X}$ from $10^{27}$ to
$10^{32}$ erg s${-1}$). As the Getman et al.\ analysis was carried out
for sources detected on the front-illuminated ACIS CCDs (I0-I3), which
have lower QE than CCD S3 at low energy (i.e., photon energies
$\stackrel{<}{\sim}0.5$ keV), application of the Getman et al.\
methods might result in an underestimate in the value of $N_H$
obtained via the median photon energy of a source on S3.  This is
unlikely to be an important effect for a source as hard as that
detected toward BP Psc in 2009, however, especially given the overall
degradation in the soft ($<1$ keV) X-ray sensitivity of ACIS in the
decade since {\it Chandra}'s
launch\footnote{http://cxc.harvard.edu/cal/memos/contam\_memo.pdf},

To determine the median detected photon energy for the BP Psc source,
we applied the
\anchor{http://www.astro.psu.edu/xray/acis/acis_analysis.html}{{\em
    ACIS Extract}} (AE) software package \footnote{The {\em ACIS
    Extract} software package and User's Guide are available at
  \url{http://www.astro.psu.edu/xray/acis/acis\_analysis.html}.}
(Broos et al. 2010) to extract and analyze the source photons. Doing
so, we find the median background-subtracted photon energy of the BP
Psc source in the 0.5--8.0 keV energy band is 3.47$\pm$0.33 keV. This
median energy corresponds to an absorbing column $\log{N_H} \, [{\rm
  cm}^{-2}] = 22.9\pm0.2$, based on the ``best model family''
determined from the Getman et al.\ (2010) statistical spectral
analysis method (we display the AE-extracted X-ray spectral energy
distribution of BP Psc, overlaid with the Getman et al.\ ``best model
family'' spectral model, in Fig.~\ref{fig:BPPsc_spec}). The Getman et
al.\ (2010) method yields an estimated apparent (measured) broad-band
flux of $F_X = (5.3\pm1.4)\times10^{-15}$ erg cm$^{-2}$ s$^{-1}$
(0.5--8.0 keV), corresponding to an intrinsic flux
$F_X=8\times10^{-14}$ erg cm$^{-2}$ s$^{-1}$. The measured and
intrinsic hard-band (2.0--8.0 keV) fluxes are
$(4.6\pm1.4)\times10^{-15}$ erg cm$^{-2}$ s$^{-1}$ and
$1.4\times10^{-14}$ erg cm$^{-2}$ s$^{-1}$, respectively.  Although
the statistical error on the broad-band intrinsic flux ($F_X$) is
small ($\sim30$\%), the value of $F_X$ is uncertain by almost a factor
of 5 (and the hard-band intrinsic flux is uncertain by a factor
$\sim$2), due to systematic uncertainties in determining the generic
spectral model that best describes the BP Psc source (see Getman et
al.\ 2010, their Fig.\ 1). However, the foregoing estimates for
apparent and intrinsic flux are consistent with those we obtained via
simple simulations to match the BP Psc source ACIS-S3 count rate using
absorbed thermal plasma models ranging in temperature from $kT_X =
1.0$ to 3.0 keV --- a range typical of coronal sources with median
photon energies similar to that of BP Psc (Feigelson et al.\ 2005;
also see \S 4) --- assuming $N_H = 10^{23}$ cm$^{-2}$.

From the preceding results for $N_H$ and $F_X$, we estimate that the
X-ray luminosity of the BP Psc source is $L_X \approx
9\times10^{28}(D/100)^2$ erg s$^{-1}$ where $D$ is the distance in
pc. Adopting a bolometric luminosity $L_{bol} \approx
8.5\times10^{33}(D/100)^2$ erg s$^{-1}$ for the star (ZMS08), and
allowing for the (large) systematic errors in $F_X$ obtained via the
Getman et al.\ (2010) method, we
conservatively conclude that $\log{(L_X/L_{bol})}$ lies in the range
$-5.8$ to $-4.2$, with a best estimate of $\log{(L_X/L_{bol})} \approx
-5.0$.
%For larger values of $N_H$, the inferred value for intrinsic
%flux increases for a given ACIS-S3 source count rate. The preceding
%upper bound on $L_X$ (and hence $L_X/L_{bol}$) would be underestimated
%by a factor $\sim3$ if $N_H$ were as large as $10^{23}$ cm$^{-2}$. We
%thus obtain stringent limits of $L_X < 3\times10^{29}(D/100)^2$ erg
%s$^{-1}$ and $\log{L_X/L_{bol}} < -4.5$ for the BP Psc X-ray source.

\subsection{Limits on diffuse X-ray emission}

As is apparent from inspection of Fig.\ 1, no diffuse X-ray emission
is associated with the jets and bow-shock-like Herbig-Haro (HH)
objects emanating from the position of BP Psc. This result is not
particularly surprising, considering such emission should be very
soft; indeed, there are few examples of shock-generated X-ray emission
among pre-MS jet/HH systems, and the handful that have been detected
to date generally display low X-ray luminosities (e.g., HH 210 in the
ONC, with observed $L_X \sim 10^{27}$ erg s$^{-1}$; Grosso et al.\
2006). We estimated an upper limit on the surface brightness of such
diffuse X-ray emission from the BP Psc jet system by counting
background events within regions centered on the H$\alpha$ emission
near the tips of the bow shocks seen in optical emission lines. The
resulting limit is $\sim10^{-15}$ erg cm$^{-2}$ s$^{-1}$
arcsec$^{-2}$. This corresponds to an upper limit on emergent X-ray
luminosity of $L_X \approx 10^{27}(D/100)^2$ erg s$^{-1}$ for any
compact ($\sim1''$) X-ray-emitting knots in the BP Psc jet system.

\subsection{Field X-ray sources}

\begin{figure}[htb]
  \centering
\includegraphics[width=6in,angle=0]{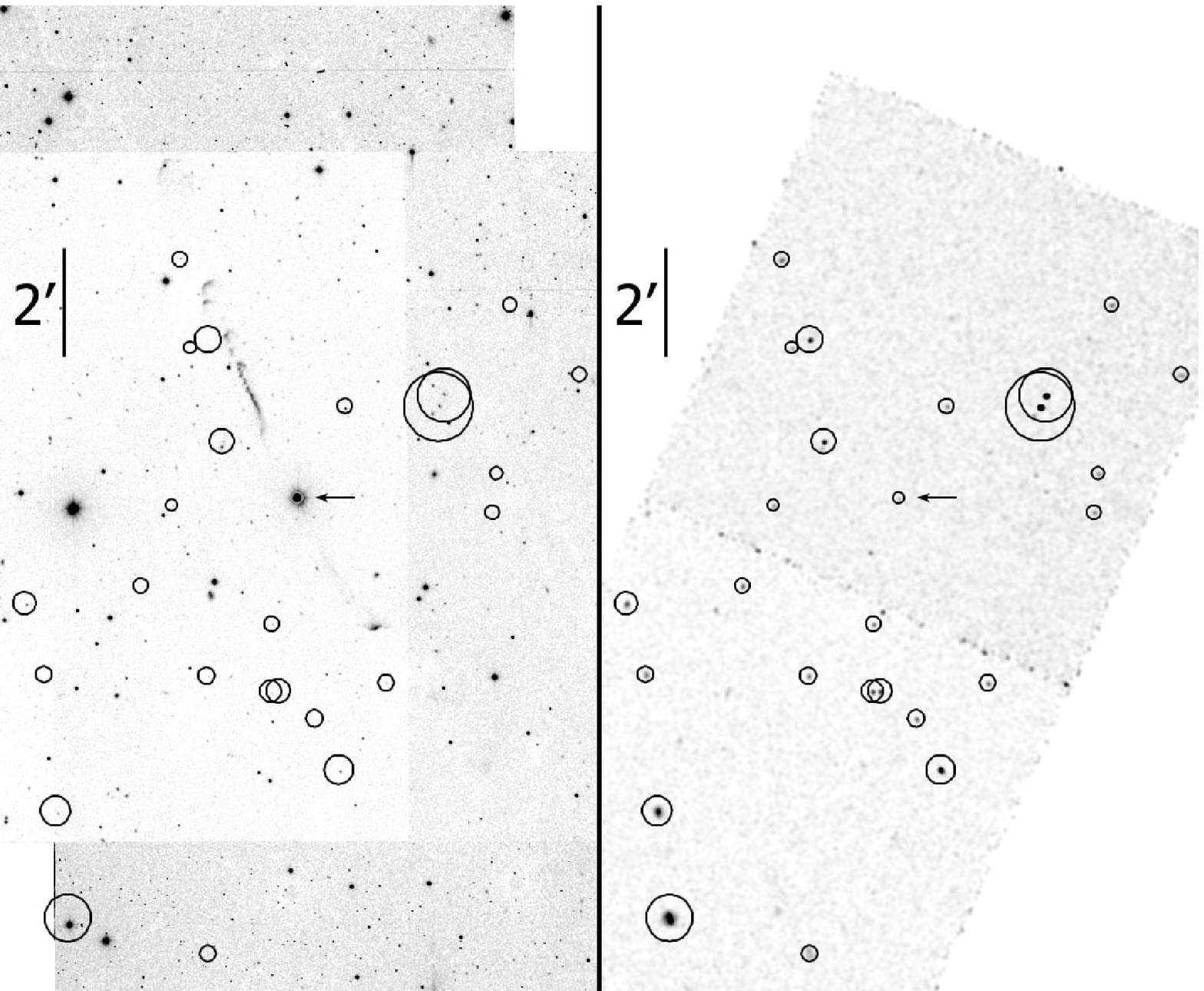}
\caption{{\it Left:} wide-field H$\alpha$ image of BP Psc and
  surrounding field, showing the jet system associated with BP Psc
  (adapted from Zuckerman et al.\ 2008). {\it Right:} broad-band
  (0.5--8.0 keV) 75.5 ks exposure {\it Chandra}/ACIS image of the same
  field. Only X-rays detected by ACIS CCDs S2 (lower region) and S3
  (upper region) are displayed; the field of view is
  $8'\times16'$. The circles on each image indicate the positions of
  {\it Chandra} X-ray sources, with radius proportional to
  signal-to-noise ratio (specifically, radii [arcsec] are set to
  2$\times$ the source signal-to-noise ratio; the smallest circles
  correspond to 3$\sigma$ detections). In each image, north is up and
  east is to the left, and the position of BP Psc is indicated by an
  arrow.}
\label{fig:BPPscImage} 
\end{figure}

If BP Psc is a nearby ($D \lesssim100$ pc), ``isolated'' classical T
Tauri star akin to TW Hya or V4046 Sgr (e.g., Kastner et al.\ 2008b
and references therein), then it could be a member of a sparse group
of young (age $\lesssim30$ Myr) stars whose members are quite X-ray
luminous relative to the field star population (Zuckerman \& Song
2004; Torres et al.\ 2008; and references therein). If so, there would
be a small chance that the $\sim0.04$ square degree field imaged by
{\it Chandra} (Fig.~\ref{fig:BPPscImage}) contains one or more
low-mass members of this putative ``BP Psc Association.''
Such young stars, being close to Earth, should be
sufficiently luminous in X-rays as to be readily detectable in our
$\sim75$ ks {\it Chandra} exposure and should be bright near-IR
sources (BP Psc itself, though occulted by its disk and not directly
detected in the near-IR, has $K \sim 7$).

As already noted, however, only one of the roughly dozen X-ray sources
detected on CCD S3 --- BP Psc itself --- is also a 2MASS source. An
additional dozen, relatively bright ($\ge 50$ count) X-ray sources are
detected on S2 (the CCD abutting S3), although some of these sources
suffer from poor image quality due to a combination of large off-axis
angle and non-optimal detector geometry with respect to the focal
surface of the {\it Chandra} mirror assembly. Of these S2 sources only
one, at $\alpha=$23:22:41.4, $\delta$=-02:21:23 (J2000) --- the bright
X-ray source at the extreme southeast corner of the image in the
righthand panel of Fig.~\ref{fig:BPPscImage} --- can be readily
associated with a bright ($V\lesssim15$) star (USNO-B1 0876--0839011).
%NOMAD1 0876--0887829). 
Given its USNO and 2MASS catalog magnitudes ($V=14.09$, $R=13.65$,
$J=12.36$, $H=11.92$, and $K=11.80$) this X-ray-bright object is
indeed a candidate young (age $\sim10$ Myr), late-type (K3-4) star;
but as such it would lie at a distance $\sim500$ pc. Furthermore, its
proper motion ($\Delta\alpha, \Delta\delta$ = [48, 28] mas yr$^{-1}$;
USNO-B1) is signicantly different from that measured for BP Psc
([44.4, $-$26.3] mas yr$^{-1}$; Tycho-2). Hence USNO-B1 0876--0839011
is not associated with BP Psc, whether or not BP Psc is a young star
at $D \lesssim100$ pc.

\section{Discussion}

\subsection{The X-ray source at BP Psc}

Our {\it Chandra} observation of BP Psc has yielded a detection of a
faint X-ray point source that is coincident with the bright
optical/near-IR and submillimeter emission associated with this
unusual field star. The nature of this X-ray emission source is
difficult to ascertain, given the small number of photons
detected. However, the median energy ($\sim3.5$ keV) and, hence, the
resulting inferred absorbing column density ($N_H \sim10^{23}$
cm$^{-2}$) appear typical of pre-MS star X-ray sources detected within
dusty circumstellar disks that are viewed nearly edge-on (Kastner et
al.\ 2005). The large value we infer for $N_H$, which corresponds to a
$V$ band extinction of $\sim100$ magnitudes (Vuong et
al.\ 2003), is consistent with the fact that the photosphere of BP Psc
is thus far only detected indirectly --- via scattering off circumstellar dust
--- in the visible and near-IR (ZMS08). 

Given that the median detected photon energy is $\sim3.5$ keV, it
would appear that the X-rays detected from BP Psc are too hard to be
attributed to internal shocks associated with its jet-launching
regions --- in contrast to cases in which soft ($<1$ keV) X-rays have
been detected at the bases of optical jets associated with certain
classical T Tauri stars (Kastner et al.\ 2005; Guedel et al.\ 2005,
2007b). Indeed, the BP Psc X-ray source spectral energy distribution
(Fig~\ref{fig:BPPsc_spec}) reveals no evidence for such a soft X-ray
``excess.'' Instead, the X-ray photons we have detected would appear
to be associated with magnetic reconnection activity near the
star. Such magnetic activity, in turn, could be due either to a
stellar corona or to star-disk interactions. In either case --- given
that the stellar photosphere is evidently obscured from direct view,
even at near-IR wavelengths (ZMS08) --- it would appear that the X-ray
source at BP Psc likely represents the first direct detection of
radiation from within a few stellar radii of BP Psc itself, and that
the X-ray-based value of $N_H$ we have obtained is therefore the only
presently available measurement of the extinction toward the star
($A_K \sim 10$).

\subsection{Constraints on a pre-main vs.\ post-main
sequence nature}

The values of X-ray luminosity and $\log{(L_X/L_{bol})}$ (\S 3.1)
inferred from the {\it Chandra} detection would appear to place BP Psc
at the extreme low end of the (well-determined) X-ray luminosity
distributions of T Tauri stars in the ONC (Preibisch \& Feigelson
2005) and in Taurus (Guedel et al. 2007a).  Specifically, the value
of $L_X$ we estimate for BP Psc, $L_X\approx 10^{29}$ erg s$^{-1}$, places
BP Psc well below the median value of $L_X$ for
$0.4-2.0 M_\odot$ ONC stars in the 3--20 Myr age range (the relevant
range if BP Psc is pre-main sequence; ZMS08). Only
$\sim2$\% of ONC stars display $L_X < 3\times10^{29}$ erg s$^{-1}$  
(Preibisch \& Feigelson 2005).

In contrast, the inferred range of $\log{(L_X/L_{bol})}$, $-5.8$ to
$-4.2$, would place BP Psc well within the (rather
broad) $\log{(L_X/L_{bol})}$ distribution of the class of rapidly
rotating (FK Com-like) G giants. The dozen or so members of this class
range from $\log{(L_X/L_{bol})} = -6.4$ to $-3.1$, with a median of
$\log{(L_X/L_{bol})} = -4.7$ (Gondoin 2005). Hence, the {\it Chandra}
identification of a weak X-ray source at the position of BP Psc
appears to throw some weight behind a post-main sequence
classification for this star --- and supports a model in which the
detected X-rays most likely arise from coronal activity, as a
consequence of its rapid rotation rate (see ZMS08). The large angular momentum
required to sustain such rapid rotation, in turn, would seem to be
consistent with the notion that BP Psc has recently consumed a
companion star (as proposed by ZMS08), and is therefore an FK Com-type
star ``in the making'' (as speculated by Kastner et al.\ 2008a). 

\section{Summary and Conclusions}

We have used the {\it Chandra} X-ray Observatory to detect a weak
X-ray point source that is coincident with the centroids of
optical/IR and submillimeter continuum emission from the
disk-enshrouded, jet-driving field star BP Psc. As the star's
photosphere is obscured throughout the visible and near-infrared, the
{\it Chandra} X-ray source appears to represent the first detection of
BP Psc itself. The {\it Chandra} X-ray image yields no detection of
the BP Psc jet system, nor evidence that BP Psc is associated with a
young stellar group or cluster.

From the median energy of the detected BP Psc source photons
($\sim3.5$ keV), we infer an absorbing column $\log{N_H} \, [{\rm
  cm}^{-2}] = 22.9\pm0.2$ and an X-ray luminosity $L_X \approx
10^{29}$ erg s$^{-1}$), with $\log{(L_X/L_{bol})}$ in the
range $-5.8$ to $-4.2$. The rather large median X-ray energy indicates
that the X-rays most likely originate with magnetic activity, as
opposed to shocks near the star, and hence are produced either in a
stellar corona or in star-disk magnetospheric interaction regions. The
inferred values of $L_X$ and $L_X/L_{bol}$ are 1--2 orders of
magnitude smaller than those typical of low-mass, pre-main sequence
stars, but are well within the range observed for rapidly-rotating (FK
Com-type) G giant stars.  Hence, the {\it Chandra} results favor an
exotic model wherein the disk/jet system of BP Psc is the result of
its very recently engulfing a companion star or giant planet, as the
primary star ascended the giant branch.

\acknowledgments{This research was supported via award number
  GO8-9004X to RIT issued by the {\it Chandra} X-ray Observatory
  Center, which is operated by the Smithsonian Astrophysical
  Observatory for and on behalf of NASA under contract NAS8--03060,
  and by NASA Astrophysics Data Analysis grant NNX09AC96G to RIT and
  UCLA. M.D.P.'s work was supported by a National Science Foundation
  Astronomy and Astrophysics Postdoctoral Fellowship. J.R.G.'s work was
  supported in part by the University of California Lab Research
  Program 09-LR-01-118057-GRAJ and NSF grant AST-0909188.}

%\begin{thebibliography}{}

\subsection*{References}

\begin{list}{}{}

\item Broos, P. S., Townsley, L. K., Feigelson, E. D., Getman,
  K. V., Bauer, F. E., \& Garmire, Gordon P. 2010, ApJ, 714, 1582

\item Feigelson, E.D., et al. 2005, ApJS, 160, 379

\item Getman, K. V., Feigelson, E. D., Broos, P. S., Townsley,
  L. K., \& Garmire, Gordon P. 2010, ApJ, 708, 1760 

\item Gondoin, P. 2005, A\&A, 444, 531

\item G\"{u}del, M., Skinner, S. L., Briggs, K. R., et al. 2005,
  ApJ, 626, L53

\item G\"{u}del, M., Briggs, K. R., Arzner, K., et al. 
  2007a, A\&A, 468, 353

\item G\"{u}del, M., Telleschi, A., Audard, M., Skinner, S. L.,
  Briggs, K. R., Palla, F., \& Dougados, C. 2007b, A\&A, 468, 515

\item Grosso, N., Feigelson, E.D.,  Getman, K.V.,  Kastner, J.H.,
  Bally, J., \& McCaughrean, M.J. 2006, A\&A, 448, L29

\item Kastner, J.H., Zuckerman, B., Weintraub, D.A., \& Forveille, T.
  D.A. 1997, Science, 277, 67 

\item Kastner, J. H., Franz, G., Grosso, N., et al. 2005, ApJS, 160, 511

\item Kastner, J. H., Zuckerman, B., \& Forveille, T. 2008a, A\&A, 486, 239

\item Kastner, J. H., Zuckerman, B., Hily-Blant, P., \&
  Forveille, T. 2008b, A\&A, 492, 469

\item Melis, C., Zuckerman, B., Song, I., Rhee, J. H., \&
  Metchev, S. 2009, ApJ, 696, 1964

\item Preibisch, T., \& Feigelson, E.D. 2005, ApJS, 160, 390

\item Press, W., et al. 1992, {\it Numerical Recipes} (Cambridge
  U. Press)

%\item Rebull, L. M., Stapelfeldt, K. R., Werner, M. W., et
%  al. 2008, ApJ, 681, 1484

\item Rodriguez, D., Kastner, J.H., Wilner, D., \& Qi, C. 2010,
  ApJ, submitted

%\item Torres, C.A.O., et al. 2006, A\&A, 460, 695

\item Torres, C.A.O., Quast, G.R., Melo, C.H.F., \& Sterzik,
  M.F. 2008, in {\it Handbook of Star Forming Regions}, ed.\ B. Reipurth,
  Astron.\ Soc.\ Pacific, San Francisco, in prep.

\item Vuong, M. H., Montmerle, T., Grosso, N., Feigelson, E. D.,
  Verstraete, L., \& Ozawa, H. 2003, A\&A, 408, 581

\item Zuckerman, B., \& Song, I. 2004, ARAA, 42, 685

\item Zuckerman, B., Melis, C., Song, I., et al. 2008, ApJ, 683, 1085 
(ZMS08)

\end{list}

%\end{thebibliography}

\end{document}